\documentclass[a4paper,11pt]{article}

\usepackage[latin1]{inputenc}

\usepackage{mathrsfs}
\usepackage{amsfonts,amssymb}
\usepackage{latexsym}
\usepackage{amsmath}
\usepackage{dsfont}
\newtheorem{teo}{Theorem}
\newtheorem{defi}{Definition}

\newtheorem{pro}{Proposition}
\newtheorem{cor}{Corollary}
\newcommand{\gtw}{\tilde{\Gamma}(W)}
\newcommand{\gtwx}{\tilde{\Gamma}(W_x)}

\newcommand{\gt}{\tilde{\Gamma}}
\newcommand{\cam}{\mathscr{F}}
\newcommand{\camlo}{\mathscr{F}(\mathcal{O})}
\newcommand{\row}{{\rho}_{{}_{W}}}
\newcommand{\rowx}{{\rho}_{{}_{W_x}}}

\newcommand{\dc}{\mathcal{O}}
\newcommand{\imp}{\tilde{\Gamma}(-\mathds{1})}

\newcommand{\cvd}{\hspace*{\fill}\nolinebreak{\hspace*\fill}$\Box$\par\vspace{4mm}}

\begin{document}

\title{Generalized Particle Statistics in Two-Dimensions: Examples from \\ the Theory of
Free Massive Dirac Field}

\author{Dario Salvitti\\
\small Dipartimento di Matematica, Università di Roma ``La
Sapienza"\\
\small P.le Aldo Moro 2, 00185 Roma, Italy\\
\small \texttt{salvitti@mat.uniroma1.it}}
\date{}
\maketitle
\begin{center}
\textit{\small Dedicated to the memory of Sabrina Picucci}

\end{center}

\begin{abstract}
\noindent In the framework of algebraic quantum field theory we
analyze the anomalous statistics exhibited by a class of
automorphisms of the observable algebra of the two-dimensional free
massive Dirac field, constructed by fermionic gauge group methods.
The violation of Haag duality, the topological peculiarity of a
two-dimensional space-time and the fact that unitary implementers do
not lie in the global field algebra account for strange behaviour of
statistics, which is no longer an intrinsic property of sectors.
Since automorphisms are not inner, we exploit asymptotic abelianness
of intertwiners in order to construct a braiding for a suitable
$C^*$-tensor subcategory of End($\mathscr{A}$). We define two
inequivalent classes of path connected bi-asymptopias, selecting
only those sets of nets which yield a true generalized statistics
operator.

\end{abstract}

\section{Introduction}
The intrinsic definition of particle statistics in the approach of
Algebraic Quantum Field Theory (AQFT) in a four-dimensional
space-time is provided by assigning to superselection sectors
equivalence classes of permutation group representations, which
describe the statistics of multiparticle states. In a
(3+1)-dimensional space-time, fields and particles obey Bose-Fermi
alternative, exhibiting the more general bosonic or fermionic
parastatistics, while in lower dimensional Minkowski space
statistics are described, in general, by braid group
representations. The first models leading to particles described by
a one-dimensional representation of the braid group
(\textit{anyons}) are in \cite{Wi}, while higher dimensional
representations describe \textit{plektons}. In a (2+1)-dimensional
space-time, strictly local quantum fields are always subject to the
normal commutation rules, but particles carrying ``topological
charges", created from the vacuum by the action of fields localized
in cones, may exhibit intermediate statistics.

The statistics of a sector describes the interchange of identical
charges. In two dimensions, DHR theory allows for two distinct
statistics operators (one the inverse of the other), since the
causal complement of a bounded region has two connected components.
The statistics operator is a topological invariant if the pairs of
spatially separated auxiliary regions can be continuously deformed
from one to the other, maintaining a relative space-like distance.
Therefore, the braid group enters in the description of DHR
superselection charges localized in two-dimensional double cones,
for intervals of the real line or in (2+1)-dimensional theories for
charges localized in space-like cones.

Superselection sectors in four-dimensional theories are classified
by equivalence classes of irreducible representations of the compact
group of internal symmetries \cite{doplicher90}. However, if the
superselection category in low dimensional theories is not symmetric
but only braided, such a group may not exist. Indeed, some models of
(1+1)-dimensional conformal fields exhibit a superselection
structure which seems not to fit any representation group theory.

Why ``generalized" particles statistics? Well, this is necessary
since the algebraic approach to local field theories which do not
fulfill Haag duality and which does not allow non inner
automorphisms of the underlying field algebra does not yield a well
defined notion of statistics. Thus, we need to extend it to physical
theories which do not fit the prescriptions of the algebraic
framework totally, as in the case of smeared-out kink operators
\cite{ruikink} in the context of the two-dimensional free massive
Dirac field in the formalism of relativistic second quantization
developed in \cite{CaRu}. There, not only is Haag duality violated
(Sect. 3), but a family of unitary operators implementing DHR
automorphisms is not in the field algebra, forcing us to explore
alternative tools. In a more general setting, field theories in
(1+1)-dimensions satisfying twisted Haag duality and the split
property for wedges and having an unbroken (i.e. unitarily
implemented) group of inner symmetries $G$ give rise to a not Haag
dual observable algebra $\mathscr{A}=\cam ^G$ \cite{muedouble}.
Split property for wedges has been proven recently by Buchholz and
Lechner for the Bose and Fermi cases \cite{lechner}. Together with
the argument in \cite{muedouble}, this proves that that the
observable algebra is not Haag dual when $\cam$ is any finite
product of free massive Bose and Fermi fields and $G$ is
non-trivial.

 We remark that ``free" anyons are studied
in a two-dimensional space-time since no (2+1)-dimensional model of
free anyons can exist\footnote{This notion of ``free" anyons refers
to the on-mass-shell nature of the Fourier transform. In d=1+1 the
anyon operators can live together in the same Hilbert space as the
free Fermions, but they are not really free in the mass-shell
sense.} \cite{mund}. In the setting of CAR algebras on the fermionic
Fock space there exists a natural notion of second quantization more
appropriate for a theory of relativistic particles. The theory of
Fermionic gauge groups \cite{CaRu,obrien} displays a wide
 class of unitarily
implementable automorphisms on the antisymmetric Fock space.

Our choice is the natural one \cite{adler}, i.e. that for which the
winding numbers (the charges) are easily computable through index
formulae \cite{rui}, while the zero charge implementers are the well
known smeared-out kink operators since they can modify the
statistics of a sector \cite{schroer}. Implementable gauge groups in
the one-particle Dirac theory lead to a model which exhibits strange
statistics. A class of Bogoliubov automorphisms unitarily
implementable in the Fock representation induce a family of
localized and transportable automorphisms of the observable algebra
\cite{adler}, implemented by non local operators which are not even
contained in the field algebra $\cam.$

 Since our investigations are
strongly influenced by the violation of Haag duality and the non
locality of implementers, which gives rise to non inner
automorphisms, we begin with a discussion of the arguments
leading to the known results, in order to emphasize the need to
clarify the notion of statistics even for theories not fulfilling
all the axioms of AQFT, and to better understand the developments
presented in this paper. Statistics of sectors, approached first
with ``classical" DHR theory
\cite{doplicher69,doplicher69II,doplicher71,doplicher74,
doplicher90,frs92}, depends not only on the charge (i.e. on the
sector), but also on a continuous parameter which indexes a
collection of unitarily implementable automorphisms which carry no
charge, but modify the statistics of the composed sector
\cite{schroer}.

Unfortunately, since the net of local observables does not fulfill
Haag duality, some results largely exploited in AQFT are no longer
true in the setting with which we are concerned here, and we are
able to produce counterexamples. However, the statistics operator
still possesses all of the formal properties as it does in DHR
theory, since unitary intertwiners between automorphisms of the same
translation equivalent class are always local elements of
$\mathscr{A},$ even if Haag duality is violated.

After computing the statistics operator formally, the question
remains as to whether the braiding obtained in this way has a
genuine meaning in terms of statistics. Actually, we cannot proceed
step by step along DHR theory alone, as it deals with local objects
and often exploits Haag duality as a fundamental technical
assumption. We are now analyzing a theory that allows for
intertwiners not lying in the algebra where endomorphisms act and
where endomorphisms are not locally inner but are inner only in a
asymptotical sense.

We appeal to a more recent notion of braiding \cite{abelianita},
where the condition of asymptotic abelianness of intertwiners allow
us to define $\textit{bi-asymptopias},$ giving rise to a braiding
for a suitable full subcategory of $\mbox{End}(\mathscr{A}).$ In
higher dimensions, Roberts has shown that a DHR sector of a non-Haag
dual net $\mathscr{A}$ extends to a DHR sector of the dual net
$\mathscr{A}^d,$ and the latter can be studied with the usual
methods \cite{palermo}. In (1+1)-dimensional massive theories this
fails since $\mathscr{A}^d,$ satisfying Haag duality and split
property for wedges, has no localized sectors as Mueger has shown in
\cite{muesuper}. The net $\mathscr{A}$ may have non-trivial
localized sectors, but they necessarily become solitons when they
are extended to $\mathscr{A}^d.$

In (1+1)-dimensional free massive Dirac field theory we exclude
those braidings that do not give rise to true statistics, since they
have their DHR counterpart in pseudo statistics operators
constructed without remaining in the same connected component.
Bi-asymptopias relative to different components are not mutually
cofinal, nor path connected, due to the geometry of a
two-dimensional space-time. A direct computation for each connected
component yields different braidings, i.e. the category is not
symmetric. Physically speaking, particles described by this theory
are neither bosons nor fermions for almost all values of the
solitonic parameter $\lambda.$

The present article is organized as follows. In Sect. 2 our
assumptions are stated: relativistic second quantization,
implementable gauge groups in the (1+1)-dimensional free massive
Dirac field theory, index formulae for smeared-out kink operators in
the formalism developed in \cite{obrien}, \cite{CaRu}, \cite{rui}.

In Sect. 3 we analyze a field theory model arising from the
fermionic gauge group theory when applied to the Dirac field in
two-dimensional Minkowski space. Some known results from
\cite{adler} are derived. Investigation of the statistics in the
framework of DHR analysis leads to anomalous behaviour of the charge
composition, and statistics is not an intrinsic property of the
sector \footnote{In the conformally invariant zero mass (or short
distance) limit the situation changes radically and the standard DHR
theory becomes again fully applicable. This phenomenon is inexorably
linked to the emergence of new sectors in this limit (the disorder
becomes charge-carrying) and has been observed in \cite{koe}}.

In Sect. 4 we prove that unitary implementers are not elements of
the global field algebra $\cam$ for almost all values of the
continuous parameter $\lambda,$ thus giving a complete
classification of their localization properties.

In Sect. 5 we prove asymptotic abelianness in order to exhibit a
pair of disjoint (i.e not path connected) bi-asymptopias which give
rise to an authentic braiding for the $C^*$-subcategory of
$\mbox{End}(\mathscr{A})$ generated by our family $\Delta$ of
automorphisms. We compute the braiding explicitly and give a natural
condition to be imposed in (1+1)-dimension in order to avoid those
braidings with no true counterpart in the DHR setting.

\section{Preliminaries}

Since we work in the algebraic setting, we briefly list the axioms
appropriate for theories where observables are defined from fields
through a principle of gauge invariance.
\begin{enumerate}
\item  The field algebra $\mathscr{F}$ is the inductive
limit of the net of von Neumann algebras $\mathcal{O} \rightarrow
\mathscr{F}(\mathcal{O})$ and its action on the Hilbert space
$\mathscr{H}$ is irreducible.
\item There exists a strongly continuous unitary representation
$L \rightarrow \mathcal{U}(L)$ of the Poincaré group $\mathscr{P}$
on $\mathscr{H}$ inducing automorphisms $\alpha_L$ of the field
algebra, and the action on local algebras is geometric, i.e.
$\alpha_L(\mathscr{F}(\mathcal{O}))= \mathscr{F}(L\mathcal{O}).$
Moreover, there exists a unit vector $\Omega \in \mathscr{H},$
\textit{the vacuum vector}, unique up to a phase, which is left
invariant by $\mathcal{U}(L), \, L \in \mathscr{P}.$ The vector
state induced by $\Omega$ is the \textit{vacuum} state $\omega_0$ of
$\mathscr{F},$
$$
\omega_0(F)=(\Omega, F\Omega).
$$
\item (\textit{Reeh-Schlieder property for double cones}) The
vacuum vector $\Omega$ is cyclic and separating for every algebra
$\mathscr{F}(\mathcal{O}).$
\item There exists a faithful
 representation $g \rightarrow \beta_g$
of a compact group $G,$ the \textit{gauge group}, by automorphisms
of $\mathscr{F}.$ $\beta_g$ commutes with  $\alpha_L$ and
$\beta_g(\mathscr{F}(\mathcal{O}))=\mathscr{F}(\mathcal{O}).$
Moreover, for $F \in \mathscr{F}(\mathcal{O}),$\, the correspondence
$g \rightarrow \beta_g(F)$ is weakly continuous.
\item (\textit{Normal commutation relations})
There exists a $k \in G,$ with $k^2=e,$ such that, setting
\begin{equation}\label{guido}
\cam_{\pm}(\dc)=\{F \in \camlo : \beta_k(F)=\pm F \},
\end{equation}
we have that $\cam_+(\dc_1)$ commutes with $\cam(\dc_2)$ and
$\cam_-(\dc_1)$ anticommutes with $\cam_-(\dc_2)$ when $\dc_1
\subset \dc_2'.$ If the unitary $\Gamma(g)$ implements the
automorphism $\beta _g,$ we can reformulate (\ref{guido}) by
requiring \textit{twisted locality}:
$$
\camlo^{\tau} \subset \cam(\dc')'
$$
where $\camlo^{\tau}:= Z\camlo Z^*,$
$Z=\frac{\mathds{1}+i\Gamma(k)}{1+i},$ defines the \textit{twisted
algebra.}
\end{enumerate}

\subsection{Relativistic second quantization}

We now fix notation and give an overview of the fundamentals of
relativistic second quantization. Let $\mathscr{H}$ be a separable
complex Hilbert space with inner product $(\; ,\, ).$ The fermionic
Fock space $\mathcal{F}_a(\mathscr{H})$ is the completion of the
vector space $ \mathscr{D}_{at}:=\bigoplus_{n=0}^{\infty}\wedge^n
\mathscr{H} $ of antisymmetric algebraic tensors with respect to the
''natural'' scalar product $ < \oplus_{n \geqslant 0}\xi_n \big|
\oplus_{n \geqslant 0}\eta_n > = \sum_{n \geqslant 0}(\xi_n,
\eta_n), $ with the standard unity vector $\Omega:=(1,0,0,...)$
defined as the \textit{vacuum vector}. For every $f\in \mathscr{H},$
the \textit{creation operator} $c^*(f)$ and its adjoint $c(f)$, the
\textit{annihilation operator}, are defined on the whole fermionic
Fock space, with
 $\|c^*(f)\|=\|f\|,$ and they satisfy the
canonical anticommutation relations:
$$
\{c(f),c(g)\}=\{c^*(f),c^*(g)\}=0
$$
$$
\{c(f),c^*(g)\}=(f,g)\mathds{1}
$$
for arbitrary $f, g \in \mathscr{H}.$  If $U \in
\mathcal{B}(\mathscr{H})$, we define the operator $\Gamma(U)$ on
$\overline{\wedge^n \mathscr{H}}$ by
\begin{equation} \label{2.6}
\Gamma(U)c^*(f_1) \cdots c^*(f_n)\Omega = c^*(Uf_1) \cdots
c^*(Uf_n)\Omega,
\end{equation}
which has the property $ \Gamma(U)\Gamma(V)=\Gamma(UV). $ In
particular, if $U\in \mathcal{U}(\mathscr{H}):=
\{A:\mathscr{H}\rightarrow \mathscr{H},\, A \mbox { unitary}\} $,
the correspondence $ c^*(f) \mapsto c^*(Uf)$ is an automorphism of
the CAR algebra, unitarily implemented by $\Gamma(U)$ (in the Fock
representation):
\begin{equation}
\Gamma(U)c^*(f)\Gamma(U)^*=c^*(Uf)
\end{equation}
as follows from (\ref{2.6}). Moreover, an arbitrary $A\in
\mathcal{B}(\mathscr{H})$ induces on
 $\mathcal{F}_a(\mathscr{H})$ a \textit{sum operator} $\mbox{d}\Gamma(A)$
such that $\exp(it\textmd{d}\Gamma (A)) = \Gamma (e^{itA}),$ which
preserves the adjoint and the commutator \cite{CaRu}.

In concrete cases, as that we discuss in this paper,
 $\mathscr{H}=\mathscr{H}_+ \oplus \mathscr{H}_-$, where
 $\mathscr{H}_{\pm}$ are copies
 of the same function space $L^2 \equiv H$.
Let $P_{\delta}$ ($\delta=+,-$) be the projectors onto
$\mathscr{H}_{\pm}$. If $A$ is an operator on $\mathscr{H},$ we set
$ A_{\delta \delta'}:= P_{\delta}AP_{\delta'},$ $
\delta,\delta'=+,-.$ In this notation, $A$ is a block matrix whose
entries are endomorphisms of $L^2$. Such a decomposition of
$\mathscr{H}$ is related to the well known fact that the free Dirac
hamiltonian $H_{m}$ has spectrum $(-\infty, -m] \cup [m, +\infty)$,
where $m \geqslant 0 $ denotes the rest mass of the particle. Here,
$P_{\pm}$ are the spectral projectors of $H_m$ onto $[m, +\infty)$
and $(-\infty, -m]$ respectively. Instead of non-relativistic second
quantization $A\mapsto \mbox{d}\Gamma(A),$ we work with another
irreducible representation of the CAR algebra, defined by
\begin{equation}\label{2.39}
\tilde{c}(f):= c(P_+f)+c^*(\overline{P_-f}).
\end{equation}
The one-particle Hilbert space is then defined by $ \mathscr{H}_1:=
P_+H \oplus \overline{P_-H} $ while the physical Hilbert space is
$\mathcal{F}_a (\mathscr{H}_1).$
 If $U\in
\mathcal{U}(\mathscr{H})$ satisfies $[U, P_{\delta}]=0$, then the
automorphism
 $\tilde{c}(f)\mapsto\tilde{c}(Uf)$
is unitarily implemented by
\begin{equation}
\widetilde{\Gamma}(U):= \Gamma(U_{++})\Gamma(\bar{U}_{--}),
\end{equation}
with the compact notation $\Gamma(U_{++})\equiv \Gamma(U_{++}\oplus
\mathds{1}), \Gamma({\bar{U}}_{--})\equiv
\Gamma(\mathds{1}\oplus{\bar{U}}_{--}) $. Here the bar over an
operator stands for the action by a fixed conjugation $J$ on
$\mathscr{H},$ i.e. $\bar{U}=JUJ.$  For arbitrary $U\in
\mathcal{U}(\mathscr{H}),$ the Shale-Stinespring theorem states that
there exists a unitary
 $\widetilde{\Gamma}(U)$ on the antisymmetric Fock space such that
$$
\tilde{c}(Uf)=\widetilde{\Gamma}(U)\tilde{c}(f)\widetilde{\Gamma}(U)^*
\qquad \forall f\in \mathscr{H}
$$
if and only if the off-diagonal parts of $U,$ namely $U_{\delta
-\delta}$, are Hilbert-Schmidt ($HS$) operators. Unitaries on
$\mathscr{H}$ inducing automorphisms of the CAR algebra which are
unitarily implementable on $\mathcal{F}_a(\mathscr{H}) $ form a
group, denoted $\mathcal{G}_2$. Let $\mathfrak{g}_2:= \{A \in
\mathcal{B}(\mathscr{H)}: A_{\delta, -\delta} \in HS \}$ be the
complex Lie algebra of $\mathcal{G}_2.$  By a suitable choice for
the phases of the unitary operator implementing the automorphism of
the CAR algebra, it is always possible to define a one-parameter
strongly continuous group
$$
\widetilde{\Gamma}(e^{itA})= e^{it \textmd{d}\tilde{\Gamma}(A)},
\quad  A=A^* \in \mathfrak{g}_2
$$
where $\mbox{d}\tilde{\Gamma}(A)$ is the self-adjoint generator. The
arbitrary additive constant in the definition of
$\mbox{d}\tilde{\Gamma}(A)$ is fixed by requiring that $(\Omega,
\mbox{d}\tilde{\Gamma}(A)\Omega)=0.$ In Sect. 3, where we deal with
the Dirac field, we shall employ the more common notation
$\pi(\phi(f))$ rather than $\tilde{c}(f).$
\begin{defi} The \emph{charge operator} Q is the generator of the
one-parameter group induced by the identity:
$$
Q:= \emph{d}\widetilde{\Gamma}(\mathds{1})= \emph{d}\Gamma(P_+)-
\emph{d}\Gamma(P_-).
$$

\end{defi}
Under the action of charge operator, fermionic Fock space splits
into a direct sum of charge sectors $ \mathcal{F}_a (\mathscr{H})=
\bigoplus_{n \in \mathds{Z}}\mathcal{F}_n,$ and
$$
\widetilde{\Gamma}(U)\mathcal{F}_n = \mathcal{F}_{n+q(U)}, \qquad U
\in \mathcal{G} _2,
$$
where $q(U)\in \mathds{Z}$ is the Fredholm index of $U_{--}.$

The additive property $ q(U_1U_2)=q(U_1)+q(U_2)$ also holds. In
other terms, if $q(U)=1,$ the vacuum sector $\mathcal{F}_0$ can be
connected to the $n$ charge sector by applying
$\widetilde{\Gamma}(U)^n,$ while $ q(e^{itA})=0, \quad  A=A^* \in
\mathfrak{g}_2, $ since $q(e^{itA})=q(\mathds{1})=0$ by virtue of
the continuity in $t$. Hence the charged sectors are left invariant
by $\widetilde{\Gamma}(e^{itA})$. We end this overview with an
identity of great relevance for computations:
$$
\Gamma(-\mathds{1})\widetilde{\Gamma}(U)=(-1)^{q(U)}\widetilde{\Gamma}(U)\Gamma(-\mathds{1}).
$$
In view of the subsequent applications, we cite three useful
propositions which establish the commutation rules between unitary
implementers and their self-adjoint generators \cite{CaRu}.

\begin{pro} \label{commutazione1}
For every $A, B \in \mathfrak{g}_2$, on the domain $\mathscr{D}$ of
finite particle vectors there holds
\begin{equation}\label{2.84}
[\emph{d}\widetilde{\Gamma}(A),
\emph{d}\widetilde{\Gamma}(B)]=\emph{d}\tilde{\Gamma}([A,B])+
C(A,B)\mathds{1},
\end{equation}
where $C(A,B):= \emph{Tr }(A_{-+}B_{+-}-B_{-+}A_{+-})$ is the
\emph{Schwinger term}.
\end{pro}
Here $\mathscr{D}:=\{F \in \mathcal{F}_a(\mathscr{H}) : F=P_l F, \,
\mbox{ for some }  l \in \mathds{N}\}$ and $P_l$ denotes the
spectral projector of the particle number operator on $[0,l].$
\begin{pro} \label{commutazione2}
Let $A, B \in \mathfrak{g}_2,\, A=A^*,\,  B=B^* \mbox{ and }
[A,B]=0$. We have:
\begin{equation} \label{2.89}
\widetilde{\Gamma}(e^{iA})\widetilde{\Gamma}(e^{iB})=
e^{-C(A,B)/2}\widetilde{\Gamma}(e^{i(A+B)}).
\end{equation}
\end{pro}

\noindent The third proposition establishes the commutation rule
between second quantization operators in the case that one of them
is a \textsl{charge shift} (i.e. it carries a non zero charge).
\begin{pro} \label{cociclo}
Let $U \in \mathcal{G}_2, \, A=A^* \in \mathfrak{g}_2, \, [U,A]=0.$
Then
\begin{equation} \label{eqn:cociclo}
\widetilde{\Gamma}(e^{iA})\widetilde{\Gamma}(U)=
e^{i(\tilde{\Gamma}(U)\Omega,\,
\emph{d}\tilde{\Gamma}(A)\tilde{\Gamma}(U)\Omega)}
\widetilde{\Gamma}(U)\widetilde{\Gamma}(e^{iA}).
\end{equation}
\end{pro}

\subsection{Implementable gauge groups in the one-particle Dirac theory}

In the theory of (1+1)-dimensional free massive Dirac field, gauge
transformations are operators of multiplication by unitary matrices
on $\check{\mathscr{H}}\equiv L^2(\mathds{R},dx)^{{\otimes}2},$ the
image of $\mathscr{H}\equiv L^2(\mathds{R},dp)^{{\otimes}2}$ by
means of the Fourier transformation $\mathcal{F},$ employed to
diagonalize the differential operator representing the free Dirac
Hamiltonian. Since we are interested in lifting these unitaries to
the Fock space, we must consider only gauge transformations which
define unitaries in $\mathcal{G}_2.$  We denote by $H_1
(\mathds{R})$ the Sobolev space, which consists of all absolutely
continuous functions of $L^2(\mathds{R})$ with derivatives in $L^2
(\mathds{R}).$ Once we have introduced the group (under pointwise
multiplication)
$$
L_e\mbox{U}(1):=\{u \in \mbox{Map}(\mathds{R},\mbox{U}(1))\, | \,
u(\cdot)-1 \in H_1(\mathds{R})\},
$$
we define two faithful unitary representations $\check{\pi}_{\pm}$
of $L_e\mbox{U}(1)$ on $\check{\mathscr{H}},$ given by
$\check{\pi}_+(u)= u(x)\oplus \mathds{1}$ and $\check{\pi}_-(u)=
\mathds{1}\oplus u(-x),$ i.e. they act as multiplication by a
function on one component space only. Then $\pi_s(u) \in
\mathcal{G}_2,$ and  two projective unitary representations
$\widetilde{\Gamma}(\pi_{\pm})$ on $\mathcal{F}_a(\mathscr{H})$ are
automatically defined.

\noindent Another global gauge transformation has the form $e^{i
\varphi_+} \oplus e^{i \varphi_-}, \varphi_{\pm}\in (0,2\pi ).$

\noindent In the case of interest to us, i.e. rest mass of the
particle $m>0,$ we put $\varphi_+=\varphi_-,$ otherwise the $HS$
condition would be violated.

We end this paragraph with a formula for $\mbox{Ind}U_{--}.$ We are
interested in operators of the form
$$
(\check{U}f)(x)=u(x)f(x), \quad u(x)\in \mbox{U}(2), \quad f\in
\check{\mathscr{H}},
$$
where the $2 \times2$ matrix $u(x)$ is assumed to be diagonal:
\begin{equation}\label{matrice}
u(x)= \left(
\begin{array}{cc}
u_+(x) & 0 \\ 0 & u_-(x)
\end{array}
\right), \quad u_{\pm}(x)\in \mathds{C}.
\end{equation}
We set $A:=\mathcal{F}^{-1}\check{A} \mathcal{F}$ for $\check{A}$ an
operator on $\check{\mathscr{H}}.$  We consider continuous
multipliers of the form (\ref{matrice}) such that, for each $s,$
there exists $u_{\infty}\in C(\{\pm 1\},\mbox{U}(1))$ satisfying
$$
u_s(x)-u_{\infty}\Big(\frac{x}{|x|}\Big)=o(1), \quad |x| \rightarrow
\infty, \quad s=+,-.
$$
These multipliers form a group, denoted by $G_h,$ and their Fredholm
indices are easily computable in view of the following \cite{rui}.
\begin{teo} \label{wind}
Let $U \in G_h.$ Then
\begin{equation}
\emph{ Ind }U_{--}= \emph{w}(u_+u_-^{-1}),
\end{equation}
where $\emph{w}$ is the winding number which, by convention, is
positive on the map $x \rightarrow \frac{x-i}{x+i}.$
\end{teo}
\vspace{1mm} In order to complete the discussion of the operators we
shall employ in the next section, we remark that all our unitaries
induce automorphisms of the CAR algebra which are unitarily
implementable on the Fock space. Indeed, if $x_1 \rightarrow
\alpha(x_1)$ is an odd, monotonously increasing, $C^{\infty}$ real
valued function, which equals $1$ at the right of the interval
$(-1,1),$ we introduce the \textsl{smeared-out} kink operators
\begin{equation}
 \check{U}_{\lambda, \epsilon}:= e^{i \pi \lambda \alpha( \cdot /\epsilon)},
\quad \lambda \in \mathds{C}, \quad \epsilon >0.
\end{equation}
The off-diagonal parts of $U_{\lambda, \epsilon}$ are $HS$ for every
$\lambda \in \mathds{C},$ so it induces Bogoliubov transformations
unitarily implementable for every  $\lambda \in \mathds{R}$
\cite{rui}.

\section{Strange statistics in two-dimensional free massive Dirac field theory}

The theory of fermionic gauge group reveals itself as a natural
setting for the construction of a model which exhibits anomalous
statistics \cite{adler}. The initial Hilbert space is
$L^2(\mathds{R},\mathds{C}^2)$. Denoting by $\mathcal{K}$ the set of
double cones in the (1+1)--dimensional Minkowski space, let
$B_{\mathcal{O}}$ be the base at time $t$ of $\mathcal{O} \in
\mathcal{K}.$ The algebra of fields localizable in $\mathcal{O}$ is
defined in the usual way,
$$
\mathscr{F}(\mathcal{O})=
\{\pi(\phi(e^{itH_m}f)):\mbox{supp}(f)\subset B_{\mathcal{O}}\}'',
$$
and the global field algebra $ \displaystyle \mathscr{F}$ is the
$C^*$-inductive limit of the net $\{\camlo\}_{\dc \in \mathcal{K}}.$
 The
gauge invariant parts (i.e. the subalgebras left invariant by
$\mbox{Ad}\gt(e^{i \gamma}),$ $\gamma \in \mathds{R}$) form the net
of observables. Let us mention that the free massive Dirac field
theory in (1+1)-dimensions fulfills twisted duality
$\mathscr{F}(\mathcal{O})^{\tau}=\mathscr{F}(\mathcal{O}')'$ (the
proof of this is independent of the space-time dimension; see
\cite{Lledo} for details), but the net of observables does not
fulfil Haag duality for double cones. Indeed, if $\mathcal{O},
\mathcal{O}_1, \mathcal{O}_2$ are double cones, with bases at $t=0,$
such that $ \mathcal{O}_1, \mathcal{O}_2$ lie in different connected
components of $\mathcal{O}',$ and if $f_i$ are test functions with $
\mbox{supp}(f_i) \subset \mathcal{O}_i \,\, (i=1,2),$ then the
observable $\pi (\phi(f_1))^* \pi(\phi(f_2))$ is contained in
$\mathscr{A}(\mathcal{O})'$ but not in $\mathscr{A}(\mathcal{O'})''$
\cite{adler,muedouble}.

The automorphism of $\mathscr{F}$ defined by
${\rho}_{{}_{Z}}(\pi(\phi(f))):=
 \pi(\phi(Zf))$
 is unitarily implemented in the Fock representation when $Z$ is
one of the following unitaries of $L^2(\mathds{R}, \mathds{C}^2)$:
$$
(U(n)f)(x)=(e^{i \pi n \varepsilon(x)}f_1(x), f_2(x)), \quad n \in
\mathds{Z}
$$
$$
(V(\lambda)f)(x)=e^{i \pi \lambda \vartheta(x)}f(x), \quad \lambda
\in \mathds{R},
$$
which correspond in (\ref{matrice}) to the choice $u_+(x)=\exp({i
\pi n \varepsilon(x)}),$ $ u_-\equiv 0$ and $u_{\pm}(x)=\exp({i \pi
\lambda \vartheta(x)}),$ respectively. Here, the functions
$\varepsilon$ and
 $\vartheta$ are characterized by the same properties of $\alpha$
relative to generic intervals $(\--\epsilon, \epsilon),$ resp.
$(\--\theta, \theta),$ instead of $(-1,1).$ We emphasize that this
result is valid only in the massive case \cite{pressley}. The gauge
group U$(1)$ acts on $\cam$ through $e^{i \gamma}\mapsto
\mbox{Ad}\gt(e^{i \gamma}),$ and the self-adjoint unitary operator
inducing the twisting is $\gt(\mathds{-1}).$

We refine Propositions \ref{commutazione1} and \ref{commutazione2}
to suit our purpose \cite{pressley}. The computation of the
statistics operator needs commutation rules between implementers
when translated to mutually space-like regions. Let $\mathcal{O}$ be
a double cone with basis $(-\epsilon, \epsilon)$ at $t=0,$ and $x,
x' \in \mathds{R}$ such that $\mathcal{O}+x'
 \subset (\mathcal{O}+x)'.$ Since the Schwinger term  for the
 pair
$V(\lambda)_x,V(\lambda ')_{x'}$ vanishes when $\mathcal{O}+x'
 \subset (\mathcal{O}+x)',$
Proposition \ref{commutazione2} establishes that the projective
representation $\gt$ is multiplicative in almost all cases of
interest to us. Proposition \ref{cociclo} applies to
 the case $U=U(n)_x$ and $e^{i A}=V(\lambda)_{x'},$ giving
$$\gt(V(\lambda)_{x'})\gt(U(n)_x) =e^{i \pi n \lambda \,
\textmd{sgn}(x-x')}\gt(U(n)_x)\gt(V(\lambda)_{x'}),$$
$$\gt(U(n)_x) \gt(U(n')_{x'})= \gt(U(n')_{x'}) \gt(U(n)_x). $$
As will be clear later, we consider only even charge $n,$ while the
real number $\lambda$ is left arbitrary. Since $\mathcal{G}_2$ is a
group, the product $U(n)V(\lambda)$ is unitarily implementable too.
We set $ W \equiv W(n,\lambda):=U(n)V(\lambda). $ Choosing the same
generating function $\vartheta=\varepsilon$ we obtain a unitary
operator on $L^2(\mathds{R}, \mathds{C}^2)$ defined by:
$$
W:= e^{i \pi (n+ \lambda)\varepsilon(\cdot)} \oplus e^{i \pi \lambda
\varepsilon(\cdot)}.
$$
Here $W$ acts on both components of $L^2(\mathds{R},\mathds{C}^2)$
as multiplication by two distinct functions.
 We note that $U(n)\equiv
W(n,0)$ and $V(\lambda) \equiv W(0,\lambda)$. In order to determine
the charge carried by the automorphism ${\rho}_{{}_W}$ we must
evaluate $q(W).$ The Fredholm index of $W_{--}$ can be easily
computed as an immediate application of Theorem \ref{wind}, and it
equals $n$. Note that
 $V(\lambda)$ belongs to the connected component of the
identity, therefore
 $\mbox{Ind} V(\lambda)=\mbox{Ind}(\mathds{1})=0$, $ \lambda \in \mathds{R}$.
Thus $n$ is the charge carried by $W,$ with no contribution from
$V(\lambda).$ (The charge is entirely transported by $U(n)$ while
$V(\lambda)$ is neutral).
\begin{pro}\label{automorfismi}
Automorphisms of $\mathscr{F}$ of the form ${\rho}_{{}_W}$ induce,
on restriction, automorphisms of the observable algebra
$\mathscr{A}$.
\end{pro}
{\it Proof.}  Since any $*$-homomorphism between $C^*$ algebras is
continuous, the statement is an immediate consequence of the
inclusion
$$ {\rho}_{{}_W}(\mathscr{A}(\mathcal{O}_1))
\subset \mathscr{A}(\mathcal{O}_1), \qquad \forall \mathcal{O}_1 \in
\mathcal{K}.
$$
In order to prove this, let us consider a unitary $W(n, \lambda)$
with generating function $\varepsilon$ centred in $\mathcal{O} \in
\mathcal{K}.$ (We assume that all double cones have base at $t=0$).
We note that
 ${\rho}_{{}_W}(\mathscr{F}(\mathcal{O}_1)) \subset  \mathscr{F}(\mathcal{O}_1).$
Indeed, if $\mbox{supp}(f) \subset B_{\mathcal{O}_1},$ then
$\mbox{supp}(Wf)\subset B_{\mathcal{O}_1}$ too, hence :
$$
{\rho}_{{}_W}(\pi(\phi(f)))\equiv \pi(\phi(Wf))\in
\mathscr{F}(\mathcal{O}_1).
$$
Moreover, $[{\rho}_{{}_W}(A), \widetilde{\Gamma}(e^{i \gamma})]=0$
for all $A\in \mathscr{A}(\mathcal{O}_1)$ and $\gamma \in
\mathds{R},$ since the adjoint actions $\mbox{Ad}\gt(e^{i\gamma})$
and $\mbox{Ad}\gt(V)$ commute between themselves in view of
Proposition \ref{cociclo}. The claim follows from the gauge
invariance of $A.$

\cvd
\begin{pro}
Automorphisms of $\mathscr{A}$ defined as in Proposition
\ref{automorfismi} are localizable in double cones.
\end{pro}
{\it Proof.} Let  $A:=\pi(\phi(f)), \, \mbox{supp}(f)\subset
B_{\mathcal{O}_1},$ where $\dc_1 \subset \dc'.$ Obviously, if $x
\notin \mbox{supp}(f)$ then $(Wf)(x)=0.$ If $x \in
\mbox{supp}(f)\subset B_{\mathcal{O}_1},$ then $x \notin
B_{\mathcal{O}}$ and $\varepsilon(x)=\pm 1,$ and we have
$$
(Wf)(x)=e^{\pm\pi i \lambda}f(x),
$$
hence ${\rho}_{{}_W}(A)=\mbox{Ad} \gt(e^{\pm  \pi i \lambda})(A), \,
\forall A \in \mathscr{F}(\mathcal{O}_1).$ Now, it is then evident
that, if $A\in \mathscr{A}(\mathcal{O}_1)\equiv
\mathscr{F}(\mathcal{O}_1)^{\textmd{U}(1)},$ then
${\rho}_{{}_W}(A)=A,$ i.e. $\row$ acts trivially on
$\mathscr{A}(\dc').$

\cvd
\begin{pro} \label{principale}
For each unitary $W$ defined as above, ${\rho}_{{}_W}:=
\emph{Ad}\widetilde{\Gamma}(W)$ defines a localizable and
translatable automorphism of $\mathscr{A}$.
\end{pro}
{\it Proof.}  We have just proved localizability: $ {\rho}_{{}_W}
|_{\mathscr{A}(\mathcal{O'})}=\mbox{id}
|_{\mathscr{A}(\mathcal{O'})}, $ where $\mathcal{O}$ is the double
cone in whose base the unitary $W$ (i.e. its generating function
$\varepsilon $) is ``centred'', and $
{\rho}_{{}_W}(\mathscr{A}(\tilde{\mathcal{O}}))\subset
(\mathscr{A}(\tilde{\mathcal{O}})) $ for every double cone
$\tilde{\mathcal{O}}\supset \mathcal{O}.$ In order to prove
translatability, let us observe that denoting the translates by
$W_x:=T(x)WT(-x)$, the automorphism ${\rho}_{{}_{W_x}}$ is localized
in $\mathcal{O}+x$. Moreover, the unitary
$\widetilde{\Gamma}(W_xW^*)$ intertwines
${\rho}_{{}_W}:=\mbox{Ad}{\tilde{\Gamma}(W)}$ and
${\rho}_{{}_{W_x}}:=\mbox{Ad}{\tilde{\Gamma}(W_x)},$ and induces an
equivalence between them since it is a local observable. Indeed,
$\widetilde{\Gamma}(W_xW^*)\in \mathscr{A}(\tilde{\mathcal{O}}),$
with $\tilde{\mathcal{O}}\supset \mathcal{O}\cup \mathcal{O}_x$. The
gauge invariance comes from the commutation relation of Proposition
\ref{cociclo} between $\widetilde{\Gamma}(V)$ and
$\widetilde{\Gamma}(e^{iA})$, where the inner product
(\ref{eqn:cociclo}) now vanishes. Indeed, since
$q(W_xW^*)=q(W_x)+q(W^*)\equiv q(W)-q(W)=0,$ it follows that the
intertwiner $\widetilde{\Gamma}(W_xW^*)$ preserves the charge and so
$\widetilde{\Gamma}(W_xW^*)\Omega \in \mathds{C}\Omega.$ We thus
have:
$$
(\widetilde{\Gamma}(W_xW^*)\Omega, \mbox{d}\widetilde{\Gamma}(\gamma
\mathds{1})\widetilde{\Gamma} (W_xW^*)\Omega)=(\Omega,
\mbox{d}\widetilde{\Gamma}(\gamma \mathds{1})\Omega)=0.
$$
 In order to prove that $\widetilde{\Gamma}(W_xW^*)\in
\mathscr{F}(\tilde{\mathcal{O}'})',$ let us consider $\mathcal{O}_1
\subset \tilde{\mathcal{O}}'$ and $\mbox{supp}(f)\subset
B_{\mathcal{O}_1}$. In order to evaluate the expression
\begin{equation} \label{c1}
\widetilde{\Gamma}(W_xW^*)\pi (\phi (f))\widetilde{\Gamma}(W_xW^*)^*
= {\rho}_{{}_{W_x}}\circ {\rho}_{{}_{W^*}}(\pi (\phi (f))),
\end{equation}
we notice that ${\rho}_{{}_{W^*}}\equiv {\rho}_{{}_W}^{-1}$ is still
localized in $\mathcal{O},$ and then, since $W(n,
\lambda)^*=W(-n,-\lambda),$
$$
{\rho}_{{}_{W^*}}(\pi (\phi (f)))=\pi(\phi(e^{-i\pi \lambda}f)).
$$
Analogously
$$
{\rho}_{{}_{W_x}}(\pi (\phi (f)))=\pi(\phi(e^{i\pi \lambda}f)),
$$
so the right side of (\ref{c1}) reduces to $\pi (\phi (f))$ and the
result follows from twisted duality.

\cvd

\noindent In the previous proof we have incidentally established
that unitaries $\gtw$ are gauge invariant if and only if $q(W)=0.$
\vspace{2mm}

 We are now in position to perform the computation of the
statistics operator $\varepsilon_{{\rho}_{{}_W}}.$ For simplicity,
we start with an automorphism ${\rho}_{{}_W}$ localized in a double
cone $\mathcal{O}$ centred at the origin.

\begin{pro} \label{statistica}
If the automorphism ${\rho}_{{}_W}$ is localized in a double cone
centred at the origin, its statistics operator is
$$
\varepsilon_{{\rho}_{{}_W}}= e^{\pm 2 \pi i n \lambda \,}\mathds{1},
$$
according to the connected component of $\dc'.$
\end{pro}
{\it Proof.} Since we work at $t=0,$ we omit the component-subscript
and consider $x \in \mathds{R}.$ The automorphism
${\rho}_{{}_{W_x}}$ is localized in
 $\mathcal{O}+x$ and is unitarily equivalent to ${\rho}_{{}_W}$ through
the intertwiner $\gt(W_xW^*)$. Then,
$$
\varepsilon_{{\rho}_{{}_W}}=\widetilde{\Gamma}(W_xW^*)^*
{\rho}_{{}_W}(\widetilde{\Gamma}(W_xW^*))=
\widetilde{\Gamma}(W)\widetilde{\Gamma}(W_x)^* \widetilde{\Gamma}(W)
\widetilde{\Gamma}(W_x)\widetilde{\Gamma}(W^*)^2.
$$
(Here, and in the sequel, we omit all cocyles since they are always
coupled with their conjugate). With this convention the previous
expression yields
$$
\widetilde{\Gamma}(W)\widetilde{\Gamma}(W_x^*)\widetilde{\Gamma}(U)\widetilde{\Gamma}(V)
\widetilde{\Gamma}(U_x)\widetilde{\Gamma}(V_x)\widetilde{\Gamma}(V)^*
\widetilde{\Gamma}(U)^* \widetilde{\Gamma}(W)^*.
$$
By Proposition \ref{cociclo} and our remarks on the specific cases
discussed in \cite{pressley},
\begin{eqnarray*}
\widetilde{\Gamma}(V) \widetilde{\Gamma}(U_x)\equiv
\widetilde{\Gamma}(e^{iX(\lambda)})\widetilde{\Gamma}(U(n)_x) \!\!&
= & \!\! e^{i(\tilde{\Gamma}(U_x)\Omega ,
\textmd{d}\tilde{\Gamma}(X(\lambda))\tilde{\Gamma}(U_x)\Omega)}
\widetilde{\Gamma}(U_x)\widetilde{\Gamma}(e^{iX(\lambda)})\\
& = & \!\!e^{i\pi n \lambda
\,\textmd{sgn}(x)}\widetilde{\Gamma}(U_x)\widetilde{\Gamma}(V),
\end{eqnarray*}
while $\widetilde{\Gamma}(V)$ commutes with
$\widetilde{\Gamma}(V_x),$ and $\widetilde{\Gamma}(U_x)^*$ with
$\widetilde{\Gamma}(U).$ We then obtain
$$
 e^{i\pi n\lambda
\,\textmd{sgn}(x)}\widetilde{\Gamma}(W)\widetilde{\Gamma}(V_x^*)
\widetilde{\Gamma}(U)\widetilde{\Gamma}(V_x)\widetilde{\Gamma}(U)^*
\widetilde{\Gamma}(W)^*.
$$
Repeating the same arguments for the two central terms, one has
$$
\varepsilon_{{\rho}_{{}_W}}=e^{2\pi i n \lambda
\,\textmd{sgn}(x)}\mathds{1}.
$$
\cvd

If the automorphism is translated to a double cone $\mathcal{O}+x,$
for arbitrary $x,$ it assumes the form ${\rho}_{{}_{W_x}}$, with $W$
localized around the origin. With a proof identical to that of Prop.
\ref{statistica}, one easily obtains the following.

\begin{pro} \label{operatore statistico}For every $x \in \mathds{R}$,
$$
\varepsilon_{{\rho}_{{}_{W_x}}}= e^{2\pi in \lambda \emph{
sgn}(x-y)}\mathds{1},
$$
with $\mathcal{O}+y$ the auxiliary double cone, spatially separated
from $\mathcal{O}+x$, used in the construction of the statistics
operator.
\end{pro}

 \noindent \textit{Remark.} In view of the decomposition
 ${\rho}_{{}_W}={\rho}_{{}_{U}}{\rho}_{{}_{V}},$ an alternative method of
evaluating $\varepsilon_{{\rho}_{{}_W}}$ is based on the identity
\begin{equation} \label{frs}
\varepsilon_{{\rho}_{{}_{U}} {\rho}_{{}_{V}}}={\rho}_{{}_{U}}
(\varepsilon({\rho}_{{}_{U}},
{\rho}_{{}_{V}}))\varepsilon_{{\rho}_{{}_{U}}} {{\rho}_{{}_{U}}}^2
(\varepsilon_{{\rho}_{{}_{V}}}){\rho}_{{}_{U}}
(\varepsilon({\rho}_{{}_{V}}, {\rho}_{{}_{U}})).
\end{equation}
A straightforward computation reduces (\ref {frs}) to $
\varepsilon_{{\rho}_{{}_W}}=
{\rho}_{{}_{U}}(\varepsilon_M({\rho}_{{}_{U}},{\rho}_{{}_{V}})),$
where the monodromy operator is simply $
\varepsilon_M({\rho}_{{}_{U}}, {\rho}_{{}_{V}})=e^{2\pi i n \lambda
\,\textmd{sgn}(y-x)}\mathds{1}.$ \\ We also observe that we could
have determined the latter by exploiting the low-dimensional quantum
field theory as formulated in \cite{frs92}, where only the
statistics phases are involved. Indeed, since ${\kappa}_{{}_{W}}=
e^{2\pi i n \lambda \,\textmd{sgn}(y-x)}$ and
${\kappa}_{{}_{U}}={\kappa}_{{}_{V}}=1,$ the claim follows from
$$
\varepsilon({\rho}_{{}_{V}},{\rho}_{{}_{U}})\varepsilon({\rho}_{{}_{U}},{\rho}_{{}_{V}})=
\frac{{\kappa}_{{}_{W}}}{{\kappa}_{{}_{U}}{\kappa}_{{}_{V}}}
\mathds{1}
$$
\cite[Lemma 3.3]{frs92}. Therefore, these results are still
consistent with the general theory of local quantum fields in low
dimension.

We have incidentally noticed that in a (1+1)-dimensional massive QFT
the statistics of a product may not coincide with the product of
statistics, i.e. composition of DHR morphisms with ordinary
statistics may generate braid statistics. This possibility is
excluded by (3+1)-dimensional QFT with Haag duality \cite[pag.
179]{doplicher69II}, where $\varepsilon_{\xi_1}\varepsilon_{\xi_2}=
\varepsilon_{\xi_1 \xi_2}$ for two arbitrary superselection sectors
$\xi_1, \xi_2,$ (i.e. equivalence classes of Poincaré covariant
localized automorphisms). The factorization property of statistics
is no longer true in theories where non ordinary statistics can
occur. In our model this property is equivalent to the triviality of
monodromy, i.e. if and only if the automorphism carries ordinary
statistics. Since unitary intertwiners between translation
equivalent automorphisms are local observables, the violation of the
multiplicative property of statistics cannot be attributed to the
violation of Haag duality but to the geometry of space-time.

It is easily seen that the statistics operator depends on the
translation equivalent class of automorphisms but not on its
representative.
  The following Corollary is then
evident. \mathstrut
\begin{cor}\, The statistics operator $\varepsilon_{{\rho}_{{}_W}}$ gives rise to
a one-dimensional representation of the braid group if and only if
$\, 2 n \lambda \notin \mathds{Z}.$
\end{cor}
We end this section with an expression for the statistics operator
$\varepsilon({\rho}_{{}_{W}},{\rho}_{{}_{W'}})$ when the unitaries
$W$ and $W'$ are centred in the same interval.  If the induced
automorphisms are localized in $\mathcal{O},$ let $x,y \in
\mathds{R}^2$ be such that $\mathcal{O}_x $ and $\mathcal{O}_y$ lie
in the right component of $\mathcal{O}',$ with $\mathcal{O}_x \succ
\mathcal{O}_y,$ where by $\mathcal{O}_x \succ \mathcal{O}_y$
 we mean that $\mathcal{O}_x$ lies in the
right component of the space-like complement of $\mathcal{O}_y$. Let
${\rho}_{{}_{W_x}}$ be localized in $\mathcal{O}_x$ and equivalent
to ${\rho}_{{}_{W}}.$ Analogously, let ${\rho}_{{}_{W'_y}}$ be
localized in $\mathcal{O}_y$ and equivalent to ${\rho}_{{}_{W'}}.$
One then has
\begin{eqnarray*}
\varepsilon({\rho}_{{}_{W}},{\rho}_{{}_{W'}})& = &
\widetilde{\Gamma}(W'{W'_y}^*) \times \widetilde{\Gamma}
(WW_x^*)\circ \widetilde{\Gamma}(W_xW^*)\times \widetilde{\Gamma}(W'_yW'^*)\\
& = &
\widetilde{\Gamma}(W'{W'_y}^*)\widetilde{\Gamma}(W'_y)\widetilde{\Gamma}(W{W_x}^*)
\widetilde{\Gamma}(W'_y)^* \widetilde{\Gamma}(W_xW^*){\rho}_{{}_{W}} (\widetilde{\Gamma}(W'_yW'^*))\\
& = &
\widetilde{\Gamma}(W'{W'_y}^*)\widetilde{\Gamma}(W'_y)\widetilde{\Gamma}(W{W_x}^*)
\widetilde{\Gamma}(W'_y)^* \widetilde{\Gamma}(W'_yW'^*)\widetilde{\Gamma}(W_xW^*)\\
& =  & e^{- i \pi (n \lambda' +n'
\lambda)}\widetilde{\Gamma}(W')\widetilde{\Gamma}(W)
\widetilde{\Gamma}(W')^* \widetilde{\Gamma}(W)^*\\
& = & e^{- \pi i (n \lambda ' + n' \lambda)}\mathds{1}.
\end{eqnarray*}
On the other hand, if we choose $\mathcal{O}_x \prec \mathcal{O}_y,$
the exponent in the last member changes sign. Therefore
\begin{equation}\label{coppia}
\varepsilon({\rho}_{{}_{W}},{\rho}_{{}_{W'}})= e^{i \pi  (n
\lambda'+ n' \lambda)\,\textmd{sgn}(y-x)}\mathds{1}.
\end{equation}

\section{Charge implementers are not quasilocal}

As already stated, charge implementers $\gtw$ do not belong to the
observable algebra. In this section we will show that they are not
even in the field algebra $\cam$ when $\lambda \neq 0 \bmod 2.$ As a
first step we observe that if $\gtw \in \cam,$ then its translates
$\gtw_x \in \cam $ too, through $
\alpha_x(\pi(\phi(f)))=\pi(\phi(\tau_x f)), $ where $(\tau_x
f)(\xi):=f(\xi -x).$  Once we have determined the statistics
operator, we can exclude the trivial case $\lambda=0$ (no kinks
present), since it yields ordinary statistics. Obviously, this is
not the unique value of $\lambda$ for which the statistics reduce to
the ordinary one. The other values which realize this possibility
depend on the charge, since they are given by $2n\lambda \in
\mathbb{Z},$ and are trivially taken into account.

For $x \in \mathbb R $ such that $\mathcal{O}+x \subset
\mathcal{O}',$ we have
\begin{equation} \label{jill}
 \gtw \gtwx = e^{2\pi i q
\lambda  \textmd{ sgn}(x)}\gtwx \gtw,
\end{equation}
where $q \equiv q(W)=q(W_x)$ is the charge carried by both $\row$
and $ \rowx.$ For a general field $F$ we have
 \begin{equation} \label{twist}
 F^{\tau}= F_+ -iF_-\imp,
\end{equation}
 where $F_{\pm}$ denote the bosonic, resp. fermionic, part of
$F.$ This can be easily seen from the explicit form of the twisted
field in the general case:
$$
F^{\tau} = \mbox{Ad}Z (F), \qquad Z=\frac{\mathds{1}+ i \gt(-
\mathds{1})}{1+i}
$$
(the symbol $\mathds{1}$ always denotes the identity operator on the
corresponding Hilbert space, as is clear from the context). We are
interested in the case $2 q \lambda \not \in \mathbb{Z},$ since this
leads to a contradiction -- the commutators are non vanishing,
according to (\ref{jill}). Let $\gtw \in \cam.$ There exists a
sequence of local fields $\{F_n\}_{n \in \mathds{N}}$ norm
convergent to $\gtw,$ with $F_n \in \cam(\mathcal{O}_n), n \in
\mathds{N}.$ Since $\gtw^{\tau}=\gtw,$ we have
\begin{equation}\label{asim}
\|F_n ^{\tau}-  F_n\| \leqslant 2 \|\gtw - F_n\|.
\end{equation}
On the other hand, according to (\ref{twist}), we have $F_n
^{\tau}-F_n= -(1+i)F_n^- Z,$ and thus $\|F_n ^{\tau}-F_n\|=
\sqrt{2}\, \|F_n^-\|.$ By virtue of (\ref{asim}) we then have
\begin{equation} \label{three}
\|F_n^-\|  \leqslant \sqrt{2} \| \gtw -F_n\|.
\end{equation}
Let now $M > 0$ be such that $\|F_n\|< M.$  An $\varepsilon /3$
argument yields
\begin{equation} \label{four}
\big \|[\gtw, \gtw_x] \big \| < (M+1) \|\gtw -F_n\| + \|F_n F_{n,x}-
\gtw_x \gtw \|,
\end{equation}
where we have used the fact that the translations, being implemented
by unitary operators, are norm-preserving. Here $F_{n,x}$ denotes
the translate of $F_n$ by $x,$ for the generic $n \in \mathds{N}.$
The second term on the right hand side of (\ref{four}) is dominated
by
$$
\|F_n  F_{n,x}- F_{n,x}F_n \| + \|F_{n,x}F_n - \gtw_x \gtw \|.
$$
Since $F_n$ and $F_{n,x}$ are local fields, $[F_n, F_{n,x}]= 2 F_n^-
F_{n,x}^-$ when $\dc_n + x \subset \dc_n '.$ In this case we obtain,
using (\ref{three}):
$$
\|F_n  F_{n,x}- F_{n,x}F_n \| \leqslant 2 \|F_n^- \| \|{F_{n,x}}^-
\| \leqslant 4\|\gtw -F_n \|^2.
$$
In the last step we have exploited the commutativity between the
actions $\alpha_{\mathds{R}^2}$ and $\beta_{\textmd{U}(1)}$ in order
to yield ${F_x}^- ={F^-}_x$ for each quasilocal field $F.$

 Let now $\varepsilon >0,$ and let $n_{\varepsilon}
\in \mathds{N}$ be a positive integer such that
$$
\|\gtw-F_n\| < \frac{\varepsilon}{3(M+1)},  \quad \|F_{n,x}F_n
-\gtw_x \gtw \| <\frac{\varepsilon}{3}
$$
for all $n \geqslant n_{\varepsilon}.$  The integer
$n_{\varepsilon}$ is independent of $x$. For $n=n_{\varepsilon},$
let $x>0$ be such that $\dc_{n_{\varepsilon}}+x \subset
{\dc_{n_{\varepsilon}}}'.$ Then, the left hand side of (\ref{four})
can be made arbitrarily small, contradicting (\ref{jill}).

\vspace{2mm}

This proof, though intuitive, does not work if $2 q\lambda \in
\mathds{Z},$ since no contradiction is obtained in the latter case.
In particular, our arguments exclude the case $q=0.$ We will give an
alternative proof which overcomes this impediment, based on twisted
duality for the field algebra.
\begin{teo} For every $\lambda \neq0\ \bmod 2$ the unitary implementers $\gtw$
are not elements of the field algebra $\mathscr{F}.$ If $\lambda =0
\bmod 2,$ they are local elements of the field algebra.
\end{teo}
{\it Proof.}  If $\widetilde{\Gamma}(W)\in \mathscr{F},\,$  let
 $\{F_n\}_{n \in \mathds{N}}$ be a sequence of local elements
of $\mathscr{F}$ norm convergent to $\widetilde{\Gamma}(W),$ with
$F_n \in \mathscr{F}(\mathcal{O}_n)$ for suitable double cones
$\dc_n.$ Since $F^{\tau}_n \in \mathscr{F}(\mathcal{O}_n)^{\tau}=
{\mathscr{F}({\mathcal{O}_n}')}',$  $ [F^{\tau}_n, A]=0, \, A\in
\mathscr{F}({\mathcal{O}_n}'), \, n \in \mathds{N}. $ If
$\hat{\mathcal{O}}_n \subset {\mathcal{O}_n}'$, we have $
[F^{\tau}_n, \pi(\phi(f))]=0, \, \mbox{ supp}(f)\subset
B_{\hat{\mathcal{O}}_n}, n \in \mathds{N}. $ With this notation it
follows that
$$
\|\widetilde{\Gamma}(W)^{\tau}\pi(\phi(f))-
\pi(\phi(f))\widetilde{\Gamma}(W)^{\tau}\| \leqslant
$$
$$
\leqslant \|(\widetilde{\Gamma}(W)^{\tau}-F_n^{\tau})\pi(\phi(f))\|+
\|F_n^{\tau}\pi(\phi(f))- \pi(\phi(f))F_n^{\tau}\|+
$$
$$
+ \|\pi(\phi(f))(F_n^{\tau}-\widetilde{\Gamma}(W)^{\tau})\|
\leqslant
$$
\begin{equation}\label{michele}
\leqslant 2\|\pi(\phi(f))\|\,
\|\widetilde{\Gamma}(W)^{\tau}-F_n^{\tau}\|+
\|F_n^{\tau}\pi(\phi(f))-\pi(\phi(f))F_n^{\tau}\|.
\end{equation}
Without lost of generality, we consider only normalized functions.
Then, since the correspondence $f \rightarrow \pi_P(\phi(f))$ is
isometric,  $\| \pi(\phi(f))\|=1.$ Fixing an arbitrary $\epsilon>0,$
let $n_{\epsilon}\in \mathds{N}$ be such that $
\|\widetilde{\Gamma}(W)^{\tau}-F_n^{\tau}\|< \epsilon/2 $ for every
$n \geqslant n_{\epsilon}.$ For $n=n_{\epsilon}$ and
$\mbox{supp}(f)\subset B_{\hat{\mathcal{O}}_{n_{\epsilon}}},$ where
$\hat{\mathcal{O}}_ {n_{\epsilon}}\subset
{\mathcal{O}_{n_{\epsilon}}}',$ the expression in the last line of
(\ref{michele}) is less than $\epsilon.$ Therefore, for each fixed
$q$ and $\lambda$, $\|\widetilde{\Gamma}(W)^{\tau}\pi(\phi(f))-
\pi(\phi(f))\widetilde{\Gamma}(W)^{\tau}\|$ can be made arbitrary
small by choosing a suitable function.

On the other hand, being
$\widetilde{\Gamma}(W)^{\tau}=\widetilde{\Gamma}(W),$ the expression
in the first line of (\ref{michele}) assumes the simpler form
\begin{equation}\label{ticco}
\| e^{-i \pi
\lambda}\pi(\phi(f))\widetilde{\Gamma}(W)-\pi(\phi(f))\widetilde{\Gamma}(W)\|=
|e^{-i \pi \lambda}-1|,
\end{equation}
in view of the CAR relations and the unitarity of
$\widetilde{\Gamma}(W).$ Thus, we have a contradiction between
(\ref{michele}) and (\ref{ticco}), since the latter establishes
that, for $\lambda \neq 0 \, \bmod 2,$ the norm is a strictly
positive constant. Finally, if $\lambda=0 \bmod 2$ the unitary
implementers are local elements of $\mathscr{F}.$ Indeed, if $W$ is
centred in $\mathcal{O},$
$$
\gtw \pi(\phi(f))\gtw^*=\pi(\phi(f))
$$
for every $f$ with support spatially separated from $\mathcal{O}.$
Therefore, by twisted duality, $ \gtw\in
\mathscr{F}(\mathcal{O}')'=\mathscr{F}(\mathcal{O})^{\tau}. $ On the
other hand, since the twisting is involutive on any local field
algebra, i.e. $\mathscr{F}(\mathcal{O})^{\tau
\tau}=\mathscr{F}(\mathcal{O}),$ $\gtw\in \mathscr{F}(\mathcal{O})$
follows from $\gtw^{\tau}\equiv \gtw.$

\cvd

This result shows that the strange behaviour of statistics for this
model appears only when the implementers are not elements of the
field algebra, confirming that there is no contradiction between
what we expected from the general theory of superselections sectors
and the peculiarities arising from this model. When the ``solitonic"
parameter $\lambda$ vanishes, the statistics is again trivial, i.e.
conventional Bose-Einstein or Fermi-Dirac. Since only the zero
charge $\gtw$ are gauge invariant, the unique cases in which the
implementers are observables (and local) are when $\lambda \in 2
\mathbb Z$ and $q=0.$ The classification of the localizability
property of our implementers is thus complete.

\section{Braiding Structure and Asymptotic Abelianness}

In order to approach the study of strange statistics with a more
general tool, we observe that the well known AQFT as formulated in
\cite{doplicher71}, \cite{doplicher74} and \cite{doplicher90} cannot
be applied here in its entirety, since Haag duality is violated by
(1+1)-dimensional free massive Dirac field theory. A more
appropriate setting seems to be that proposed in \cite{abelianita}
in order to construct symmetric tensor $C^*$-categories in QED,
since it extends to theories where intertwiners are not contained in
the algebra where the endomorphisms act. Moreover, the endomorphisms
are not necessary locally inner \footnote{We recall that
$\widetilde{\Gamma}(W)\notin \mathscr{A}$ in almost all cases.}, but
only in an asymptotic sense. Asymptotic abelianness yields a tensor
$C^*$-category starting directly from representations, without
exploiting Haag duality. We set up notation and state definitions.

Let us introduce two sets of nets ${\rho}_{{}_{W}} \mapsto
\mathcal{U}_{{\rho}_{{}_{W}}}$ and ${\rho}_{{}_{W}} \mapsto
\mathcal{V}_{{\rho}_{{}_{W}}}$ for every object ${\rho}_{{}_{W}}$.
Each net consists of unitary intertwiners in
$({\rho}_{{}_{W}},{\rho}_{{}_{W_{x_m}}})$, where
${\rho}_{{}_{W_{x_m}}}$ tends pointwise in norm to the identity
morphism on $\mathscr{A}$ for suitable sequences $\{x_m\}_{m}.$
\begin{defi}[Asymptotic abelianness] \label{def:abelianita}
A field theory model satisfies asymptotic abelianness if, given
intertwiners $R\in ({\rho}_{{}_{W}}, {\rho}_{{}_{W'}}),$ \,$S \in (
{\rho}_{{}_{\widetilde{W}}},{\rho}_{{}_{\widetilde{W}'}})$ and nets
$U_m \in \mathcal{U}_{{\rho}_{{}_{W}}}, \, U'_{m'}\in
\mathcal{U}_{{\rho}_{{}_{W'}}}, V_n \in \mathcal{V}_{
{\rho}_{{}_{\widetilde{W}}}},$ $ V'_{n'} \in
\mathcal{V}_{{\rho}_{{}_{\widetilde{W}'}}}$,
\begin{equation} \label{eqn:abelianita}
U'_{m'}RU^*_m \times V'_{n'}SV^*_n - V'_{n'}SV^*_n \times
U'_{m'}RU^*_m \longrightarrow 0
\end{equation}
in norm as $ m,m',n,n' \rightarrow \infty $.
\end{defi}

Finally, the sets of nets must be compatible with products: for each
pair $\,{\rho}_{{}_{W}}, {\rho}_{{}_{W'}} \in \Delta, \,$ there
exist $U_m \in \mathcal{U}_{{\rho}_{{}_{W}}} \mbox{ and } U_{m'}\in
\mathcal{U}_{{\rho}_{{}_{W'}}} \mbox{ such that } U_m \times
U_{m'}\in \mathcal{U}_{{\rho}_{{}_{W}} {\rho}_{{}_{W'}}},$ and
similarly for $\mathcal{V}.$ Here $\Delta$ denotes a semigroup of
endomorphism of $\mathscr{A}.$ If the nets satisfy all these
conditions, they give rise to a \textit{bi-asymptopia}.
\begin{teo}\label{teo:abelianita} \emph{\cite{abelianita}}
If ${\rho}_{{}_{W}},  {\rho}_{{}_{\widetilde{W}}}\in \Delta $, then:
$$
\varepsilon({\rho}_{{}_{W}}, {\rho}_{{}_{\widetilde{W}}}):=
\lim_{m,n \rightarrow \infty} V^*_n \times U^*_m \,\,\,  U_m \times
V_n
$$
exists, is independent of  $U_m \in \mathcal{U}_{{\rho}_{{}_{W}}}$
and $V_n \in \mathcal{V}_{ {\rho}_{{}_{\widetilde{W}}}},$ and lies
in $({\rho}_{{}_{W}}  {\rho}_{{}_{\widetilde{W}}},
 {\rho}_{{}_{\widetilde{W}}}{\rho}_{{}_{W}}).$ Moreover, if $R \in
({\rho}_{{}_{W}},{\rho}_{{}_{W'}}),$ \, $S \in (
{\rho}_{{}_{\widetilde{W}}},{\rho}_{{}_{\widetilde{W}'}})$ and
${\rho}_{{}_{\widehat{W}}}\in \Delta,$
 then:
$$
\varepsilon({\rho}_{{}_{W'}},{\rho}_{{}_{\widetilde{W}'}})\circ R
\times S = S \times R \circ \varepsilon({\rho}_{{}_{W}},
 {\rho}_{{}_{\widetilde{W}}}),
$$
$$ \varepsilon({\rho}_{{}_{W}}  {\rho}_{{}_{\widetilde{W}}},
{\rho}_{{}_{\widehat{W}}})= \varepsilon({\rho}_{{}_{W}},
{\rho}_{{}_{\widehat{W}}}) \times \mathds{1}_{
{\rho}_{{}_{\widetilde{W}}}} \circ \mathds{1}_{{\rho}_{{}_{W}}}
\times \varepsilon(
{\rho}_{{}_{\widetilde{W}}},{\rho}_{{}_{\widehat{W}}}),
$$
$$
\varepsilon({\rho}_{{}_{W}},
{\rho}_{{}_{\widetilde{W}}}{\rho}_{{}_{\widehat{W}}})= \mathds{1}_{
{\rho}_{{}_{\widetilde{W}}}} \times \varepsilon({\rho}_{{}_{W}},
{\rho}_{{}_{\widehat{W}}}) \circ \varepsilon({\rho}_{{}_{W}},
 {\rho}_{{}_{\widetilde{W}}}) \times \mathds{1}_{{\rho}_{{}_{\widehat{W}}}}.
$$
\end{teo}
\vspace{2mm}

\noindent We apply this method to our class $\Delta$ of automorphism
in the setting of the (1+1)-dimensional free massive Dirac field. In
light of the computation performed in the previous section, it
remains to verify the condition of asymptotic abelianness in order
to have a bi-asymptopia and consequently a braiding for our
category. Starting from (2+1)-dimensional models, where a space-like
cone $\mathcal{C}$ and its opposite $\mathcal{-C}$ are usually
chosen as asymptotic localization regions in the definition of the
families
 $\mathcal{U}$ and $\mathcal{V},$ we extends the method to a
 (1+1)-dimensional space-time by choosing the two standard wedges
$W_{\pm}.$
 Since in (1+1)-dimensions there is a natural notion of right and left,
hence of $+\infty$ and $-\infty$, we set
$$
\mathcal{U}_{{\rho}_{{}_{W}}}:=\{(U_a)_a \subset
({\rho}_{{}_{W}},{\rho}_{{}_{W_a}}), \,\, a \rightarrow +\infty\},$$
$$
\mathcal{V}_{{\rho}_{{}_{W}}}:=\{(V_b)_b \subset
({\rho}_{{}_{W}},{\rho}_{{}_{W_b}}), \,\, b \rightarrow -\infty\},$$
where the two families of nets are contained in $W_+$, $W_-$
respectively. (This condition implies that the nets are contained in
the causal complement of every bounded region for large values of
the indexes). Let $R: {\rho}_{{}_{W}} \rightarrow {\rho}_{{}_{W'}},$
$S:
 {\rho}_{{}_{\widetilde{W}}}\rightarrow {\rho}_{{}_{\widetilde{W}'}},$ and $z,\zeta
\in \mathds{C}$ be defined by
$$
R= z \,\widetilde{\Gamma}(W'W^*), \quad S= \zeta
\,\widetilde{\Gamma}(\widetilde{W}'\widetilde{W}^*).
$$
As intertwiners $U_m \in \mathcal{U}_{{\rho}_{{}_{W}}},\, U'_{m'}\in
\mathcal{U}_{{\rho}_{{}_{W'}}}, \, V_n \in \mathcal{V}_{
{\rho}_{{}_{\widetilde{W}}}}, \, V'_{n'} \in
\mathcal{V}_{{\rho}_{{}_{\widetilde{W}'}}},$ we set:
\begin{eqnarray}
U_m & = & \lambda_m \widetilde{\Gamma}(W_{x_m}W^*), \hspace{8mm}
     x_m \hspace{1mm} \stackrel{W_+}{\longrightarrow} + \infty \nonumber \\
U'_{m'} & = & \lambda'_{m'}
\widetilde{\Gamma}(W'_{x'_{m'}}W'^*),\quad \hspace{1.5mm}
     x'_{m'} \stackrel{W_+}{\longrightarrow} + \infty \nonumber \\
V_n & = & \mu_n
\widetilde{\Gamma}(\widetilde{W}_{y_n}\widetilde{W}^*),\quad
\hspace{5.6mm}
     y_n  \hspace{2.2mm} \stackrel{W_-}{\longrightarrow} - \infty \nonumber \\
V'_{n'} & = & \mu'_{n'}
\widetilde{\Gamma}(\widetilde{W}'_{y'_{n'}}\widetilde{W}'^*),\quad
\hspace{2.4mm}
        y'_{n'} \hspace{2mm} \stackrel{W_-}{\longrightarrow} - \infty \nonumber
\end{eqnarray}
where $\lambda_m, \lambda'_{m'}, \mu_n, \mu'_{n'} \in \mathds{C}$
are defined as the scalar $z,$  for every $m,m',n,n'.$  To simplify
notation, in the sequel $x'_m$ stands for $x'_{m'}$. The term of the
sequence to which we refer will be clear from the context. Analogous
simplifications will be adopted for $y'_{n'}, \, \lambda'_{m'}$
$\mu'_{n'}.$ We now perform the computation of the limit in
(\ref{eqn:abelianita}), which now assumes the form:

$$
\lambda'_m \widetilde{\Gamma}(W'_{x'_m}W'^*)z
\widetilde{\Gamma}(W'W^*)\bar{\lambda}_m
\widetilde{\Gamma}(W_{x_m}W^*)^* \times
$$
$$
\times \mu'_n
\widetilde{\Gamma}(\widetilde{W}'_{y'_n}\widetilde{W}'^*)\zeta
\widetilde{\Gamma}(\widetilde{W}' \widetilde{W}^*) \bar{\mu}_n
\widetilde{\Gamma}(\widetilde{W}_{y_n}\widetilde{W}^*)^*-
$$
$$
- \mu'_n
\widetilde{\Gamma}(\widetilde{W}'_{y'_n}\widetilde{W}'^*)\zeta
\widetilde{\Gamma}(\widetilde{W}' \widetilde{W}^*) \bar{\mu}_n
\widetilde{\Gamma}(\widetilde{W}_{y_n}\widetilde{W}^*)^* \times
$$
$$
\times \lambda'_m \widetilde{\Gamma}(W'_{x'_m}W'^*)z
\widetilde{\Gamma}(W'W^*)\bar{\lambda}_m
\widetilde{\Gamma}(W_{x_m}W^*)^*=
$$
$$
=\lambda'_m \bar{\lambda}_m z \, D(W'_{x'_m}W'^*,W'W^*)
D(W'_{x'_m}W^*,WW^*_{x_m})
\widetilde{\Gamma}(W'_{x'_m}W^*_{x_m})\times
$$
\begin{equation} \label{d1}
\times\mu'_n\bar{\mu}_n\zeta \,
D(\widetilde{W}'_{y'_n}\widetilde{W}'^*,\widetilde{W}'
\widetilde{W}^*)
D(\widetilde{W}'_{y'_n}\widetilde{W}^*,\widetilde{W}\widetilde{W}^*_{y_n})\widetilde{\Gamma}
(\widetilde{W}'_{y'_n}\widetilde{W}^*_{y_n})- (\leftrightarrow),
\end{equation}

\mathstrut \noindent where ($\leftrightarrow$) denotes the cross
product of the same terms in the inverse order. Here $D(A,B)$ is the
cocycle of the projective representation $\widetilde{\Gamma}$, i.e.
$\widetilde{\Gamma}(W_1)\widetilde{\Gamma}(W_2)=D(W_1, W_2)
\widetilde{\Gamma}(W_1W_2).$ Collecting all scalars in a factor $q$,
(\ref{d1}) reduces to
\begin{equation} \label{d2}
q \, \Big(\widetilde{\Gamma}(W'_{x'_m}W^*_{x_m})\times
\widetilde{\Gamma}(\widetilde{W}'_{y'_n}\widetilde{W}^*_{y_n})-
\widetilde{\Gamma}(\widetilde{W}'_{y'_n}\widetilde{W}^*_{y_n})\times
 \widetilde{\Gamma}(W'_{x'_m}W^*_{x_m})\Big)
\end{equation}
so, to evaluate the asymptotic behaviour of (\ref{eqn:abelianita})
it suffices to compute the limit of the expression in parentheses.
Since
$$
\widetilde{\Gamma}(W'_{x'_m}W^*_{x_m}):
{\rho}_{{}_{W_{x_m}}}\!\!\longrightarrow {{\rho}_{{}_{W_{x'_m}'}},
\quad
\widetilde{\Gamma}(\widetilde{W}'_{y'_n}{\widetilde{W}^*}_{y_n}):
{\rho}_{{}_{\widetilde{W}_{y_n}}}\!\! \longrightarrow
{\rho}_{{}_{\widetilde{W}_{y'_n}'}}}
$$
the expression in (\ref{d2}) becomes:
$$
q \Big(
\widetilde{\Gamma}(W'_{x'_m}W^*_{x_m})\widetilde{\Gamma}(W_{x_m})
\widetilde{\Gamma}(\widetilde{W}'_{y'_n}\widetilde{W}^*_{y_n})\widetilde{\Gamma}(W_{x_m})^*-
$$
$$
-
\widetilde{\Gamma}(\widetilde{W}'_{y'_n}\widetilde{W}^*_{y_n})\widetilde{\Gamma}
(\widetilde{W}_{y_n}) \widetilde{\Gamma}(W'_{x'_m}W^*_{x_m})
\widetilde{\Gamma}(\widetilde{W}_{y_n})^*\Big)=
$$
$$
= q' \Big(
\widetilde{\Gamma}(W'_{x'_m})\widetilde{\Gamma}(\widetilde{W}'_{y'_n})
\widetilde{\Gamma}(\widetilde{W}^*_{y_n})\widetilde{\Gamma}(W_{x_m})^*
-
\widetilde{\Gamma}(\widetilde{W}'_{y'_n})\widetilde{\Gamma}(W'_{x'_m})
\widetilde{\Gamma}(W_{x_m})^*
\widetilde{\Gamma}(\widetilde{W}^*_{y_n})\Big),
$$
where
$$
q':= q \, \,\overline{D(W'_{x'_m},W^*_{x_m})
D(\widetilde{W}'_{y'_n},\widetilde{W}^*_{y_n})}.
$$
By construction  $y'_n<x'_m,\,\, y_n<x_m$ for $m,n$ sufficiently
large, then unitaries $W'_{x'_m}$ and $\widetilde{W}'_{y'_n}$ are
centred in disjoint intervals for $m,n$ sufficiently large.
Equivalently, the corresponding implementers
$\widetilde{\Gamma}(W'_{x'_m})$ and
$\widetilde{\Gamma}(\widetilde{W}_{y'_n})$ are localized in causally
disjoint regions. We are then in a position to exploit the
commutation rules between second quantization operators we have
established in Sect. 2. For example,
$$
\widetilde{\Gamma}(W'_{x'_m})\widetilde{\Gamma}(\widetilde{W}'_{y'_n})=
 e^{-i\pi
(n\lambda'+n'\lambda)}\widetilde{\Gamma}(\widetilde{W}'_{y'_n})
\widetilde{\Gamma}(W'_{x'_m}),
$$
and analogously for $W_{x_m}$. Performing the substitutions, it
turns out that (\ref{eqn:abelianita}) vanishes for large values of
the indexes and thus, a fortiori, tends to zero. We have thus shown
the following
\begin{pro}\label{teo:miaabelianita}
The subcategory of $\emph{End}(\mathscr{A})$ generated from $\Delta$
admits a braiding structure $\varepsilon$.
\end{pro}
This model shows that the method of asymptotic abelianess, in the
form stated in \cite{abelianita}, cannot be applied to
(1+1)-dimensional massive theories in order to obtain a generalized
statistics operator, since the definition of bi-asymptopias is not
consistent with the geometric peculiarity of space-time which may
give rise to two distinct statistics operators. More precisely, we
compare the braiding
 $\varepsilon$ of Theorem
\ref{teo:abelianita} with the statistics operator
$\varepsilon({\rho}_{{}_{W}}, {\rho}_{{}_{W'}})$  computed in the
purely algebraic setting as in (\ref{coppia}). If $\{x_m\}_m$ and
$\{y_n\}_n$ are such that $ \mathcal{O}_{x_m}\succ
\mathcal{O}_{y_n}$ for sufficiently large $m$ and $n$, one has
$$
V^*_n\times U^*_m \, \, U_m \times V_n =
$$
$$
= \bar{\mu}_n\widetilde{\Gamma}
(\widetilde{W}\widetilde{W}^*_{y_n})\times \bar{\lambda}_m
\widetilde{\Gamma}(WW^*_{x_m}) \,\circ \, \lambda_m
\widetilde{\Gamma}(W_{x_m}W^*)\times \mu_n
\widetilde{\Gamma}(\widetilde{W}_{y_n}\widetilde{W}^*)=
$$
$$
\widetilde{\Gamma}(\widetilde{W}\widetilde{W}^*_{y_n})\widetilde{\Gamma}(\widetilde{W}_{y_n})
\widetilde{\Gamma}(WW^*_{x_m})\widetilde{\Gamma}(\widetilde{W}^*_{y_n})
\widetilde{\Gamma}(W_{x_m}W^*)\widetilde{\Gamma}(W)
\widetilde{\Gamma}(\widetilde{W}_{y_n}\widetilde{W}^*)\widetilde{\Gamma}(W)^*,
$$
which coincides with the expression of $\varepsilon({\rho}_{{}_{W}},
{\rho}_{{}_{\widetilde{W}}})$ gained by transporting
${\rho}_{{}_{W}}$ and $ {\rho}_{{}_{\widetilde{W}}}$ resp. to
${\rho}_{{}_{W_{x_m}}}$ and ${\rho}_{{}_{\widetilde{W}_{y_n}}}.$
Since the two nets of double cones are contained in distinct
components of $\dc',$ this is incompatible with the basic
prescriptions of AQFT, since this procedure would be equivalent to
not remaining in a fixed connected component! We remark that
asymptotic abelianness requires that the two nets of double cones
 $\mathcal{O}^{\mathcal{U}}_{m}$ and
$\mathcal{O}^{\mathcal{V}}_{n},$ which appear in
$\mathcal{U}_{{\rho}_{{}_{W}}}$ and $\mathcal{V}_{{\rho}_{{}_{W}}},$
do lie in distinct components of $\mathcal{O}'$, since only this
configuration guarantees that $\mathcal{O}^{\,\mathcal{U}}_m -
\mathcal{O}^{\mathcal{V}}_n $ tends space-like to infinity as $m,n
\rightarrow \infty .$ (There are no alternatives in (1+1)-dimension
with no additional constraint). So, the theory of bi-asymptopias may
not lead to true statistics when the Minkowski space is not at least
(2+1)-dimensional and braid statistics occurs. As in the AQFT
setting, we can always collect path connected bi-asymptopias into
equivalence classes, but without additional prescription on the
double limit in Definition \ref{def:abelianita}, we could include
objects which have no physical meaning, since they correspond in
(1+1)-dimensions to working with both components of $\dc'$ at the
same time. A natural way to generalize this approach to
bidimensional theories is to reformulate some definitions, giving a
restricted notion of asymptotic abelianness appropriate to all
dimensions. Instead of performing the double limit as in Definition
\ref{def:abelianita} and Theorem \ref{teo:abelianita}, we choose a
particular ``direction", for example the diagonal one, i.e. $m=n,$
$$
\varepsilon(\row, {\rho}_{{}_{\widetilde{W}}}):= \lim_{n \rightarrow
\infty} V^*_n \times U^*_n \,\,\,  U_n \times V_n,
$$
provided $\dc_{x_n}$ and $\dc_{y_n}$ are space-like separated for
$m$ and $n$ sufficiently large. (Analogously for the definitions of
asymptotic abelianness and bi-asymptopias). All properties continue
to be valid, since the true reason for sending to infinity the two
nets of double cones is to exploit morphisms which commute.
 For example, with the definition
$\dc_n^{\mathcal{V}}:= \dc_n^{\mathcal{U}}\pm \hat{e}_{{}_1} n, $
all conditions are satisfied and each pair of nets lies in the wedge
$W_{\pm}$ for large values of the indexes. (Here $\hat{e}_{{}_1}$
denotes the unit vector in the $x_1$-direction). In this way, the
braiding structure arising from asymptotic abelianness of
intertwiners coincides with the braiding computed for a Haag-dual
net whose morphisms are all inner, and it gives rise to a true
braided tensor $C^*$-category in all other cases, e.g. the present
one, (where the statistics operator now has a genuine, not simply
formal, meaning of statistics). In other terms, we have excluded all
cases in which intertwiners satisfy asymptotic abelianness, but the
limits in Theorem \ref{teo:abelianita} do not give rise to a
statistics operator. The braiding induced by these bi-asymptopias is
really non symmetric, since bi-asymptopias
 $\{\mathcal{U},\mathcal{V}\},$
$\{\mathcal{V},\mathcal{U}\}$ are not path connected. In (3+1)
dimensions all particles exhibit ordinary statistics since we can
choose $\mathcal{O}^{\mathcal{V}}_n= - \mathcal{O}^{\mathcal{U}}_n$
when we deal with strictly localized morphisms.  This ensures the
possibility of changing continuously from
$\{\mathcal{U},\mathcal{V}\}$ to $\{\mathcal{V},\mathcal{U}\}$ along
a chain of double cones. In (2+1) dimensions, where cones give the
better notion of localization, one can choose $\rho_a$ in such a way
that $a$ tends to space-like infinity remaining in a space-like cone
$\mathcal{C},$ resp. $-\mathcal{C},$ for $\mathcal{U}_{\rho}$,
$\mathcal{V}_{\rho},$ and it is always possible to interchange the
two cones by a sequences of allowed moves. This is not possible in
(1+1) dimensions, since $\mathcal{O}'$ is not connected and thus, a
fortiori, is not arcwise connected. We remark that a distinction
between $\varepsilon(\rho,\sigma)$ and $\varepsilon(\sigma,\rho)^*$
for a generic pair of DHR morphisms cannot be achieved by
interchanging the roles of $\mathcal{U}$ and $\mathcal{V},$ since
both nets of double cones are in the same connected component of
$\dc'.$ Hence, in two dimensions to each morphism (object) we must
associate two bi-asymptopias, one for each side of $\dc'$. A direct
computation of the limits in Theorem \ref{teo:abelianita} for each
connected component separately, gives two distinct values, $e^{\pm 2
\pi i n \lambda},$ which
coincide with those found before.\\

\noindent \textit{Remark.} \,Although two arbitrary automorphisms of
the form ${\rho}_{{}_{W}}$ are always connected by a similarity
transformation (i.e.,
${\rho}_{{}_{W_2}}=\mbox{Ad}{\widetilde{\Gamma}(W_2 W^*_1)}\circ
{\rho}_{{}_{W_1}}$), this does not imply that they are unitarily
equivalent through a local element of the observable algebra, since
the unitary intertwiner is not necessarily in $\mathscr{A}.$ More
precisely, if $(n_1,\lambda_1)\neq (n_2,\lambda_2)$, then
$\widetilde{\Gamma}(W_1W^*_2) \notin \mathscr{A}$ as a consequence
of the previous observation and of group relations for unitaries
$W(\, \cdot \, , \, \cdot \, ; \, \varepsilon),$ i.e.
$W(n,\lambda)W(n',\lambda')=W(n+n',\lambda+\lambda'),\, \,
W(n,\lambda)^*=W(-n,-\lambda).$ On the other hand, products $W_xW^*$
have a different behaviour, since
\begin{equation}\label{eqn:locale}
W(n,\lambda; \varepsilon)_x W(n,\lambda; \varepsilon)^* = W(n,
\lambda; \tau_x\varepsilon - \varepsilon)
\end{equation}
and, as already stated, $ \widetilde{\Gamma}(W_xW^*)\in
\mathscr{A}(\tilde{\mathcal{O}}), \mbox{ where }
\tilde{\mathcal{O}}\supset \mathcal{O}\cup \mathcal{O}_x. $ We
emphasize that the notation in ($\ref{eqn:locale}$) may give rise to
ambiguities, since the operator on the right hand side carries no
charge even if $n\neq 0,$ due to the particular form of the
generating function.

\vspace{3mm}

\section*{Conclusions}

In the setting of AQFT we have shown that a family of localized and
transportable automorphisms of the observable algebra $\mathscr{A}$
exhibits non ordinary statistics. Inside each sector one has
different braiding structures labelled by a solitonic parameter
$\lambda$ which reflects the action of smeared-out kink operators
carrying no charge.

Owing to non locality of charge implementers, statistics is not an
invariant of the sector, as already known in some two-dimensional
particle theories or in solitonic theories. On the underlying
ordinary structure, smeared-out kink operators give rise to a
continuous family of braided tensor categories in the sense of the
theory of bi-asymptopias.

The results are consistent with AQFT, which must be handled
carefully here when tackling problems arising from the non locality
of unitary implementers, the violation of Haag duality and the
topological peculiarity of (1+1)-dimensional space-time. Owing to
the latter, some results of local field theory are no longer valid
in a two-dimensional world, giving rise to a range of intermediate
situations and strengthening the concept that for massive theories
in (1+1) dimensions statistics is not an intrinsic characteristic of
sectors a priori \cite{schroer2}. In the present case, since Haag
duality can be overcome by peculiarities of the model, strangeness
of statistics has its origin in the fact that implementers do not
lie in the field algebra.

The interpretation of the braiding structure of this model extends
to the CAR algebra the constructive method exploited for the Weyl
algebra. Since not all the braidings obtained in this way give rise
to a notion of statistics compatible with the DHR analysis, but only
those constructed from pairs of sets of nets which tend to the
\textsl{same} space-like infinity, the method of bi-asymptopias can
be carried over to (1+1)-dimensional space-time only if we add a
compatibility condition. This kind of selection criterion reflects
the ``initial condition" which determines uniquely the statistics
operator in the standard algebraic approach, i.e. trivialization of
$\varepsilon({\rho}_{{}_{W}}, {\rho}_{{}_{W'}})$ for
${\rho}_{{}_{W'}}\prec {\rho}_{{}_{W}}$ \cite{frs92}. The particles
described by this model are ``statistical schizons", since the same
sector allows ``pseudo" statistical descriptions and they exist in
the same Hilbert space either as bosons or as fermions or as proper
anyons \cite{schroer}, i.e. in two-dimensional massive theories not
only the spin but also the statistics is a convention.
 \vspace{3mm}
\subsection*{Acknowledgement} I am greatly indebted to the supervisor
of my PhD thesis, S. Doplicher, for many helpful discussions and
much encouragement. I would like to thank C. D'Antoni, J. Roberts
and M. Gabriel for a careful reading of the manuscript, and B.
Schroer for useful correspondence. It is a pleasure to thank A.~
Silva, who was the coordinator of the PhD program at the Department
of Mathematics of ``La Sapienza", University of Rome, in the period
when this work has been carried out.

\end{document}